\documentclass[namedreferences]{solarphysics}
\usepackage[hyperref,optionalrh,solaromanenum]{spr-sola-addons} 
\usepackage{amssymb}
\usepackage{latexsym}
\usepackage{wrapfig}
\usepackage{epsfig}                     
\usepackage{graphicx}                    
\usepackage{subfig}
\usepackage{natbib}
\usepackage{color}                       
\usepackage{breakurl}                         

\newcommand{\md}{\mathrm{d}}
\newcommand{\im}{\mathrm{i}}

\usepackage{tikz}

\begin{document}

\begin{article}

\begin{opening}

\title{Standing Slow MHD Waves in Radiatively Cooling Coronal Loops}

%
\author{K.S.~\surname{Al-Ghafri}
       }

%
\runningauthor{K.S.~\surname{Al-Ghafri}}
\runningtitle{Standing Slow MHD Waves in Radiatively Cooling Coronal Loops}

%
  \institute{College of Applied Sciences - Ibri,  \\
P.O. Box: 14 Ibri, Postal Code: 516, Oman\\
                     email: \url{khalil.ibr@cas.edu.om} 
             }

\begin{abstract}
The standing slow magneto-acoustic oscillations in cooling coronal loops are investigated. There are two damping mechanisms which are considered to generate the standing acoustic modes in coronal magnetic loops namely thermal conduction and radiation. The background temperature is assumed to change temporally due to optically thin radiation. In particular, the background plasma is assumed to be radiatively cooling. The effects of cooling on longitudinal slow MHD modes is analytically evaluated by choosing a simple form of radiative function that ensures the temperature evolution of the background plasma due to radiation coincides with the observed cooling profile of coronal loops. The assumption of low-beta plasma leads to neglect the magnetic field perturbation and eventually reduces the MHD equations to a 1D system modelling longitudinal MHD oscillations in a cooling coronal loop. The cooling is assumed to occur on a characteristic time scale much larger than the oscillation period that subsequently enables using the WKB theory to study the properties of standing wave. The governing equation describing the time-dependent amplitude of waves is obtained and solved analytically. The analytically derived solutions are numerically evaluated to give further insight into the evolution of the standing acoustic waves. We find that the plasma cooling gives rise to a decrease in the amplitude of oscillations. In spite of the reduction in damping rate caused by rising the cooling, the damping scenario of slow standing MHD waves strongly increases in hot coronal loops. 
\end{abstract}


\end{opening}

\section{Introduction}
Observation made by SUMER instrument on board SoHO confirms that  longitudinal standing (slow) magneto--acoustic waves have been seen to suffer a rapid damping \citep{Kliem02,Wang02,Wang03b}. Modelling for observed oscillations by SUMER have been carried out by \cite{Taroyan07}, where excitation and damping mechanisms are investigated. It is also presented how standing and propagating slow MHD waves can be differentiated from each other. The issue of fast damping is still obscure and attracts a remarkable attentions to understand the dominant method of damping. The most nominated mechanism for damping the standing longitudinal oscillations are thermal conduction, radiation and viscosity \citep{Ofmanwang, Moortel03,Mendoza04,Taroyan05,Sigalotti07,Al-Ghafri12,Al-Ghafri14}. Radiation is found to dominate the damping of cool coronal loops whereas hot loops are damped by thermal conduction. For a recent review of standing slow waves in coronal loops see \cite{Wang11}.

Plasma cooling has been detected everywhere in the solar atmosphere \citep{viall12,viall13}. In the absence of coronal heating, the plasma starts to cool by radiation and thermal conduction \citep{Klimchuk12}. For example, \cite{Aschwanden08} have shown that the coronal loops are cooling with the characteristic cooling time of the order of a few oscillation periods. \citet{Morton09b,Morton09c,Ruderman11a,Ruderman11b} investigated the effect of cooling on kink oscillations of coronal loop. They found that the kink oscillations experience an amplification due to the cooling. Further to this, propagating slow MHD waves in a homogeneous, radiatively cooling plasma are studied by \cite{Morton10a}. Essentially, plasma cooling is reported to cause a strong damping for longitudinal slow MHD waves. Recently, \cite{Erdelyi11} investigated the behaviour of longitudinal magneto--acoustic oscillations in slowly varying coronal plasma. In particular, the damping rate is found to undergo a reduction by the emergence of cooling. 

\cite{Al-Ghafri12} and \cite{Al-Ghafri14} have studied the effect of cooling on standing slow MHD waves in hot coronal loops that are damped due to thermal conduction. In particular, the plasma cooling is assumed by unspecified thermodynamic source. The individual effect of cooling is found to amplify the amplitude of oscillating loops. Although the rate of damping caused by thermal conduction increases at the beginning, it is noticed to reduce gradually in a hot corona because of plasma cooling. As thermal conduction approaches infinity, the damping rate tends to zero. Hence, slow magnetosonic waves propagate almost without damping at the slower, isothermal sound speed.

The present article discusses the combined effects of radiative cooling and thermal conduction on damping longitudinal standing MHD waves. A function with a simple form representing the radiation mechanism is taken to ensure that the plasma temperature decreases exponentially with time according to observations. The cooling is assumed to be weak with the characteristic cooling time scale much larger than the oscillation period. The paper is structured as follows. In Section~(\ref{the model}) we present our model and derive the main governing equation with boundary conditions. The analytical results are obtained with the aid of the WKB theory in Section~(\ref{analytic sol}). In Section~(\ref{numerical}) the individual and combined effects of radiation and thermal conduction are studied by displaying the analytical solution numerically. Our discussions and conclusions are presented in Section~(\ref{conclusion}).

\section{The Model and Governing Equations}\label{the model}
We model a straight coronal loop in which the magnetic field is uniform and in the {\it z}-direction, {\it i.e.} $\mathbf{B}_0=B_0\mathbf{\hat z}$. The magnetic loop is of length $L$ with its ends at $z=\pm L/2$ . The background plasma temperature (pressure) is assumed to change as a function of time due to radiative cooling. The density is a constant.

The governing MHD equations for the background plasma take the following form
\begin{eqnarray}
&&\frac{\partial{\rho}}{\partial{t}}+\nabla.(\rho\mathbf{v}) =0,\label{Eq:cont}\\
&&\rho\frac{\partial{\mathbf{v}}}{\partial{t}}+\rho(\mathbf{v}.\nabla)\mathbf{v}=-\nabla{p}+\frac{1}{\mu_0}(\nabla\times\mathbf{B})\times\mathbf{B},\\
&&\frac{R}{\tilde\mu}\frac{\rho^\gamma}{(\gamma-1)}\left[\frac{\partial{}}{\partial{t}}\frac{T}{\rho^{\gamma-1}}
+(\mathbf{v}.\nabla)\frac{T}{\rho^{\gamma-1}}\right]=\nabla(\kappa_{\|}\nabla{T})-\rho^2Q(T) ,\\
&&{p}=\frac{R}{\tilde\mu}\rho{T},\label{Eq:gas-law}\\
&&\frac{\partial{\mathbf{B}}}{\partial{t}}=\nabla\times(\mathbf{v}\times\mathbf{B}).\label{Eq:induction}
\end{eqnarray}
Here, $\mathbf{v}$, $\mathbf{B}$, $\rho$, $p$ and $T$ represent the flow velocity, magnetic field, density, pressure and temperature respectively; $\mu_0$ is the magnetic permeability of free space, $R$ is the gas constant, $\tilde{\mu}$ is the mean molecular weight and $\gamma$ is the ratio of specific heats. The thermal conduction term is $\nabla(\kappa_{\|}\nabla{T})$ where $\kappa_{\|}=\kappa_0T^{5/2}$ and $\rho^2Q(T)$ is the general radiation term for optically thin losses. 
\bigskip

Note that the radiative loss function is approximated by a piecewise continuous function \citep{Rosner78,Priest} but it is assumed here to take a simple form to mach the observed cooling as described in the following. 
The observation shows that the radiative-cooling coronal loops are cooling exponentially \citep{Aschwanden08,Morton09b,Morton09c} and the cooling profile has the form
\begin{equation}
T=T_{0i}\exp(-\frac{t}{\tau_c}),\label{cooling_time}
\end{equation}
where $T_{0i}$ is the initial temperature at $t=0$ and $\tau_c$ is the cooling time scale.

In order to match the observed exponential cooling of the background plasma, \cite{Erdelyi11} assigned thermal conduction to be the essential cause of cooling in their model. Further to this,  an unspecified thermodynamic source for creating the plasma cooling was suggested by \cite{Al-Ghafri12} and \cite{Al-Ghafri14} to investigate longitudinal MHD Waves in dissipative time-dependent plasma. Moreover, \cite{Morton10a} assumed that the plasma is cooling radiatively and the radiative loss function has the form $\delta p$, where the loss term is assumed to follow the Newton cooling. Hence, we assume here that the radiation term $\rho^2Q(T)\sim\delta p$. Therefore the background plasma state with no background flow satisfies the equations
\begin{eqnarray}
&&v_0=0,\;\rho_0=\textrm{const}.,\;\Longrightarrow\; \nabla p_0=0,\label{Eq:back-motion}\\
&&p_0=\frac{R}{\tilde{\mu}}\rho_0T_0,\qquad {\it i.e.}\quad p_0\sim T_0,\label{Eq:back-gas_law}\\
&&\frac{R}{\tilde\mu}\rho_0\frac{\md{T_0}}{\md{t}}
=-\delta p_0, \;\Longrightarrow\; \frac{\md{T_0}}{\md{t}}=-\delta T_0,\label{Eq:back-energy}
\end{eqnarray}
where the 0 index denotes background quantity and $\delta$ is a small quantity. Equation~(\ref{Eq:back-energy}) gives the solution
\begin{equation}
T_0=T_{0i}\exp(-\delta t).\label{Eq:tem_decay}
\end{equation}
Comparing Equation~(\ref{Eq:tem_decay}) with Equation~(\ref{cooling_time}) it is obtained that $\delta=1/\tau_c$ and this gives a justification of taking $\delta$ to be small where the observed cooling times are $500\,\rm{s}<\tau_c<2000\,\rm{s}$ \citep{Morton10a}.

Now, we linearise the governing Equations~($\ref{Eq:cont}-\ref{Eq:induction}$) about the background state by writing all the variables on the form
$$
f(z,t)=f_0(t)+f_1(z,t),
$$
and neglecting nonlinear terms, where the subscript 1 indicates the perturbed quantities. Thus, the linearised MHD equations for the longitudinal motion can be reduced to 
\begin{equation}
\frac{\partial^2{v}_{1}}{\partial t^2}-\frac{\gamma p_0}{\rho_0}\frac{\partial^2{v}_{1}}{\partial z^2} =-(\gamma-1)\frac{\kappa_0 T_0^{5/2}}{\rho_0}\frac{\partial^3T_1}{\partial z^3}-\delta\frac{\partial v_1}{\partial t}.\label{Eq:energy}
\end{equation}
 It is more convenient to use the non-dimensional variables to solve the governing Equation~(\ref{Eq:energy}). Hence, we introduce the dimensionless  quantities
\begin{equation}
\tilde{t}=\frac{t}{P},\quad \tilde{z}=\frac{z}{L},\quad
\tilde{c}_s = \sqrt{\frac{T_0}{T_{0i}}}, \quad
\tilde{v}_{1} = \frac{{v}_{1}}{c_{si}}, \quad
\tilde{T}_{1}=\frac{T_{1}}{T_{0i}}, \quad
c_{si}^2 = \frac{\gamma RT_{0i}}{\tilde{\mu}},
\label{dimensionless}
\end{equation} 
where $P$ is the period of the loop oscillation, $\tilde{c}_s$ is the dimensionless sound speed, $c_{si}$ is the initial sound speed, and we put $P = L/c_{si}$\/. In what follows we drop the tilde.

Applying non-dimensionlisation for Equation~(\ref{Eq:energy}), we arrive at 
\begin{equation}
\frac{\partial^2{v}_{1}}{\partial t^2}-c_s^2\frac{\partial^2{v}_{1}}{\partial z^2} =-\frac{\sigma}{\gamma}c_s^{5}\frac{\partial^3T_1}{\partial z^3}-\epsilon\frac{\partial v_1}{\partial t}.\label{Eq:energy_dim-less}
\end{equation}

Here, the dimensionless constants $\sigma$ and $\epsilon$ represent the strength of thermal conduction and radiation and are defined by
\begin{equation}
\sigma=\frac{(\gamma-1)\tilde{\mu}\kappa_0\,T^{5/2}_{0i}}
   {RL\sqrt{\gamma\, p_{0i}\,\rho_0}}, \qquad \epsilon=P\delta.
\end{equation}
Both quantities $\sigma$ and $\epsilon$ are found to be small under standard coronal conditions in which $\gamma = 5/3$, $\tilde{\mu} \approx 0.6$, $R = 8.3\times10^3\,\textmd{m}^2\,\textmd{s}^{-2}\,\textmd{K}^{-1}$, and
$\kappa_0 \approx 10^{-11}\,\textmd{m}^2\,\textmd{s}^{-1}\,\textmd{K}^{-5/2}$\/. If we take $L = 100\;\textmd{Mm}$ and $T_{0i} = 0.6- 5\:\textmd{MK}$ as typical coronal values then we obtain $0.0068 \lesssim \sigma \lesssim 0.48$.

Now, the governing Equation~(\ref{Eq:energy_dim-less}) can be written in the form of wave-like equation as 
\begin{equation}
\frac{\partial}{\partial t}\left[\frac{1}{c^7_s}\left(\frac{\partial^2{v}_{1}}{\partial t^2}-c_s^2\frac{\partial^2{v}_{1}}{\partial z^2}\right)\right] = \frac{\sigma}{\gamma}\frac{\partial}{\partial z^2} \left[\gamma\frac{\partial}{\partial t}\left(\frac{1}{c_s^2}\frac{\partial v_1}{\partial t}\right)-\frac{\partial^2v_1}{\partial z^2}\right] - \epsilon\frac{\partial}{\partial t} \left(\frac{1}{c_s^7}\frac{\partial v_1}{\partial t}\right).\label{Eq:governing}
\end{equation}
Because we are interested in investigating the damping of standing waves so it is necessary to introduce the boundary conditions at $z=\pm1/2$. Therefore, as the loop ends embedded in the dense photosphere it is suitable to assume that the perturbed velocity vanishes at these ends,
\begin{equation}
v_1 = 0 \quad \mbox{at} \quad z = \pm 1/2 .
\label{bound_v1}
\end{equation}
In case of $\epsilon=\sigma=0$ ({\it i.e.} in the absence of radiative cooling and thermal conduction), Equation~(\ref{Eq:governing}) represents a simple wave equation with constant sound speed (see \opencite{Al-Ghafri12}). The effect of cooling and thermal conduction on longitudinal slow MHD waves will be investigated in the next sections.

\section{Analytical Solution}\label{analytic sol}
Now, we aim to solve the governing Equation~(\ref{Eq:governing}) using the WKB method. Thus, we need to introduce two slow variables $t_1=\epsilon t$ and $\sigma=\epsilon\tilde{\sigma}$, so Equation~(\ref{Eq:governing}) becomes
\begin{equation}
\frac{\partial}{\partial t_1}\left[\frac{1}{c^7_s}\left(\epsilon^2\frac{\partial^2{v}_{1}}{\partial t_1^2}-c_s^2\frac{\partial^2{v}_{1}}{\partial z^2}\right)\right] = \frac{\tilde{\sigma}}{\gamma}\frac{\partial}{\partial z^2} \left[\gamma\epsilon^2\frac{\partial}{\partial t_1}\left(\frac{1}{c_s^2}\frac{\partial v_1}{\partial t_1}\right)-\frac{\partial^2v_1}{\partial z^2}\right] - \epsilon^2\frac{\partial}{\partial t_1} \left(\frac{1}{c_s^7}\frac{\partial v_1}{\partial t_1}\right).\label{Eq:governing-slowVariable}
\end{equation}
Then, the WKB theory implies that the solution to Equation~(\ref{Eq:governing-slowVariable}) has the form 
\begin{equation}
v_1(z,t_1) = Q(z,t_1)\exp\left(\im\epsilon^{-1}\Theta(t_1)\right),
\label{Eq:wkb1}
\end{equation}
where function $Q$ is expanded in power series with respect to $\epsilon$\/, i.e.,
\begin{equation}
Q = Q_0 + \epsilon\, Q_1 + \dots .
\label{power-series}
\end{equation}
Substituting Equations~(\ref{Eq:wkb1}) and (\ref{power-series}) into Equation~(\ref{Eq:governing-slowVariable}) and taking the largest order terms in $\epsilon$ (order of $\epsilon^{-3}$) we obtain 
\begin{equation}
\frac{\partial^2Q_0}{\partial z^2}+\frac{\omega^2}{c_s^2}Q_0=0,\label{Eq:highest-order}
\end{equation}
where $\omega=\md\Theta/\md t_1$ and $Q_0$ is subject to the boundary conditions
\begin{equation}
Q_0=0 \qquad \textrm{at}\quad z=\pm\frac{1}{2},\label{Eq:condition-highest-order}
\end{equation}
according to Equation~(\ref{bound_v1}). 
The general solution to the boundary-value problem, Equations~(\ref{Eq:highest-order}) and (\ref{Eq:condition-highest-order}), can be given by the Fourier series in the form
\begin{equation}
Q_0(z,t_1)=\sum_{n=0}^{\infty} A_n(t_1)\cos\left((2n+1)\pi z\right), \qquad \omega_n=(2n+1)\pi c_s,\;n=0,1,2,\cdots.\label{Eq:Standing_radiation}
\end{equation}
In this study, we are only interested in the fundamental longitudinal mode corresponding to $n=0$. Thus, Equation~(\ref{Eq:Standing_radiation}) reduces to
\begin{equation}
Q_0(z,t_1)=A(t_1)\cos\left(\pi z\right), \qquad \omega=\pi c_s,\label{Eq:Standing_radiation-fundamental}
\end{equation}
where $A=A_0$ and $\omega=\omega_0$. Function $A(t_1)$ refers to the amplitude of standing oscillation. In order to find the function $A(t_1)$
we collect terms of the order of $\epsilon^{-2}$. Then, we obtain the equation
\begin{equation}
\frac{\partial^2Q_1}{\partial z^2}+\frac{\omega^2}{c_s^2}Q_1=\frac{\im}{c_s^2}\left[\left(\frac{9}{2}\omega+3\frac{\mathrm{d}\omega}{\mathrm{d} t_1}+ \tilde{\sigma}\omega^3 c_s^3  \right)Q_0+3\omega\frac{\partial Q_0}{\partial t_1} +\frac{5}{2}\frac{c^2_s}{\omega}\frac{\partial^2Q_0}{\partial z^2}+\frac{c^2_s}{\omega}\frac{\partial^3Q_0}{\partial t_1\partial z^2}- \frac{\tilde{\sigma}}{\gamma}\frac{c^7_s}{\omega}\frac{\partial^4Q_0}{\partial z^4} \right],\label{Eq:next-order}
\end{equation}
where the function $Q_1$ satisfies the boundary conditions
\begin{equation}
Q_1=0 \qquad \textrm{at}\quad z=\pm\frac{1}{2}.\label{Eq:condition-second-order}
\end{equation}
The Sturm-Lioville problem, Equations~(\ref{Eq:next-order}) and (\ref{Eq:condition-second-order}), has a non-trivial solution when the right-hand side of Equation~(\ref{Eq:next-order}) satisfies the compatibility condition. To obtain the compatibility condition we multiply Equation~($\ref{Eq:next-order}$) by $Q_0$, integrate with respect to $z$ over $[-1/2, 1/2]$ and use the integration by parts. Hence, we arrive at
\begin{equation}
\int_{-1/2}^{1/2}\frac{\im}{c_s^2}\left[\left(\frac{9}{2}\omega+3\frac{\mathrm{d}\omega}{\mathrm{d} t_1}+ \tilde{\sigma}\omega^3 c_s^3\right)Q_0^2+3\omega\, Q_0\frac{\partial Q_0}{\partial t_1} +\frac{5}{2}\frac{c^2_s}{\omega}Q_0\frac{\partial^2Q_0}{\partial z^2}+\frac{c^2_s}{\omega}Q_0\frac{\partial^3Q_0}{\partial t_1\partial z^2}- \frac{\tilde{\sigma}}{\gamma}\frac{c^7_s}{\omega}Q_0\frac{\partial^4Q_0}{\partial z^4}\right]\md z=0.
\label{Eq:cooling-compatibility_cond}
\end{equation}
The solution to Equation~(\ref{Eq:cooling-compatibility_cond}), with the aid of Equation~(\ref{Eq:Standing_radiation-fundamental}), gives the amplitude of standing wave in the form
\begin{equation}
A(t_1)=A(0)\exp\left(\frac{-1}{4}t_1+\frac{\tilde{\sigma}}{5}(\frac{\gamma-1}{\gamma})\pi^2\,[c_s^5(t_1)-1]\right).
\end{equation}
This Equation in terms of the scaled variables can be written as 
\begin{equation}
A(t)=A(0)\exp\left(\frac{-\epsilon}{4}t+\frac{\sigma}{5\epsilon}\left(\frac{\gamma-1}{\gamma}\right)\pi^2\,[c_s^5(t)-1]\right),\label{Eq:wave-amplitude}
\end{equation}
where $c_s^5(t)=\exp(-5\epsilon t/2)$. In case of $\epsilon=0$ it can be rid of the quantity $\epsilon$ ,in the denominator of second term in the exponent, using Taylor expansion for the sound speed.
  
\section{Numerical Evaluations}\label{numerical}
In this section, the analytical solution describing the temporal evolution of longitudinal standing-mode amplitude are studied using numerical evaluations. Typical coronal conditions are exploited to calculate the wave amplitude numerically. Then, the results are depicted to show the behaviour of MHD waves in radiatively cooling coronal loops.
\subsection{The effect of radiative cooling}
\begin{figure}[!ht]
\centering
\includegraphics[height=0.45\textheight,width=0.6\textwidth]{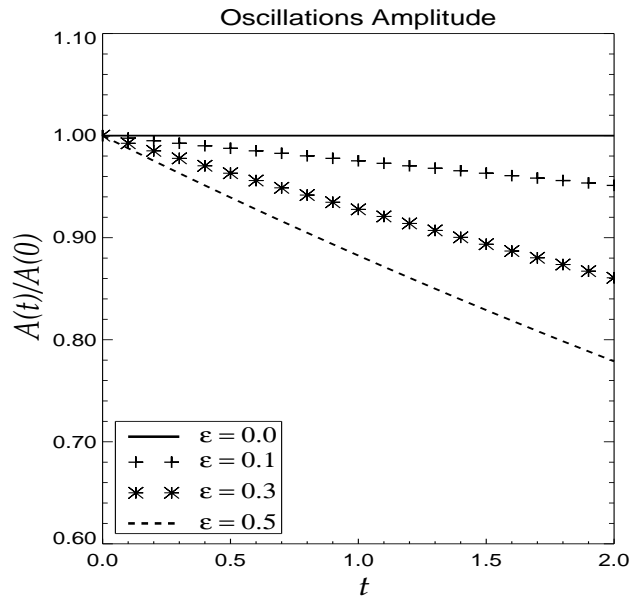}
\vspace{-0.8cm}
\caption{The amplitude of the standing wave with different values of $\epsilon$ $(0.0, 0.1, 0.3, 0.5)$ representing the effect of the cooling on the amplitude of standing wave. The time is measured in units of $L/c_{si}$.}\label{cooling-effect}
\end{figure}

In the absence of thermal conduction ($\sigma=0$) Equation~(\ref{Eq:wave-amplitude}) reduces to 
\begin{equation}
A(t)=A(0)\exp\left(\frac{-\epsilon}{4}t\right).\label{Eq:Amplitude-cooling}
\end{equation}
It is clear that from Equation~(\ref{Eq:Amplitude-cooling}) the amplitude of oscillating loops decreases with time due to radiative cooling. To give more insight  into amplitude variations we take $\epsilon\in[0, 0.5]$ as typical values for solar corona. Figure~\ref{cooling-effect} shows that the cooling causes a strong damping for the coronal loops. This result is applicable to TRACE loops with temperature $T_0=1-2$ MK where radiation is the dominant damping mechanism. 

In contrast with the result obtained by \cite{Al-Ghafri12}, who assumed the cooling by unspecified thermodynamic source, the cooling by radiation brings about an attenuation in the amplitude of waves where a strong cooling ($\epsilon=0.5$) leads to a strong damping.
\subsection{The effect of thermal conduction}
In the absence of cooling ($\epsilon=0$) Equation~(\ref{Eq:wave-amplitude}), after using Taylor expansion for the sound speed, becomes
\begin{equation}
A(t)=A(0)\exp\left(-\frac{\sigma}{2}\left(\frac{\gamma-1}{\gamma}\right)\pi^2\,t\right).\label{Eq:amplitude-thermal}
\end{equation}
This expression is consistent with its counterpart in \cite{Al-Ghafri12}. The amplitude of oscillation is damped and the strength of damping depends on the value of thermal conduction, $\sigma$. Taking into account that thermal conduction, $\sigma$, is calculated in an order of $\epsilon$ to obtain the amplitude expression, so Equation~(\ref{Eq:amplitude-thermal}) is physically applicable for a small $\sigma$. This means that Equation~(\ref{Eq:amplitude-thermal}) determines the influence of weak thermal conduction, $\sigma\ll1$. The variations of amplitude are studied on initial temperatures $T_0=[0.6, 1, 3, 5]\time10^6$ K which correspond to $\sigma=[0.0068, 0.019, 0.17, 0.48]$. 

\begin{figure}[!ht]
\centering
\includegraphics[height=0.45\textheight,width=0.6\textwidth]{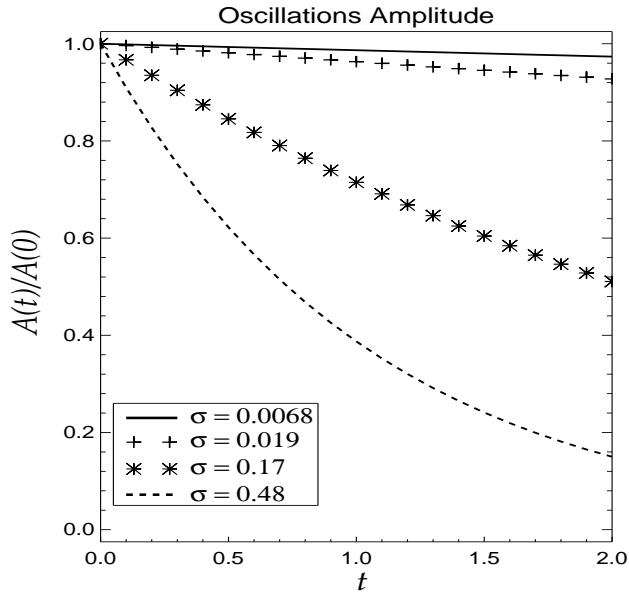}
\vspace{-0.8cm}
\caption{The amplitude of the standing wave with different values of $\sigma$ $(0.0068, 0.019, 0.17, 0.48)$ representing the effect of thermal conduction on the amplitude of standing wave. The time is measured in units of $L/c_{si}$.}\label{thermal-effect}
\end{figure}

In Figure~\ref{thermal-effect} we present the effect of varying the magnitude of the thermal conduction, $\sigma$, on the damping rate of the standing acoustic wave. As we can see that the increase of thermal conduction gives rise to a strong decline in the amplitude of standing oscillations due to the presence of thermal conduction. This result is mostly expected because thermal conduction is the essential cause of damping for the observed hot coronal loops, especially in the region of temperature $T_0\ge3$ MK. 
\subsection{Combined effects of radiative cooling and thermal conduction}
Now, we investigate the combined effects of radiation and thermal conduction on damping the amplitude of standing slow magneto-acoustic oscillations in radiatively cooling coronal loops using Equation~(\ref{Eq:wave-amplitude}). 
 \begin{figure}    
   \centerline{\hspace*{0.015\textwidth}
               \includegraphics[width=0.515\textwidth,clip=]{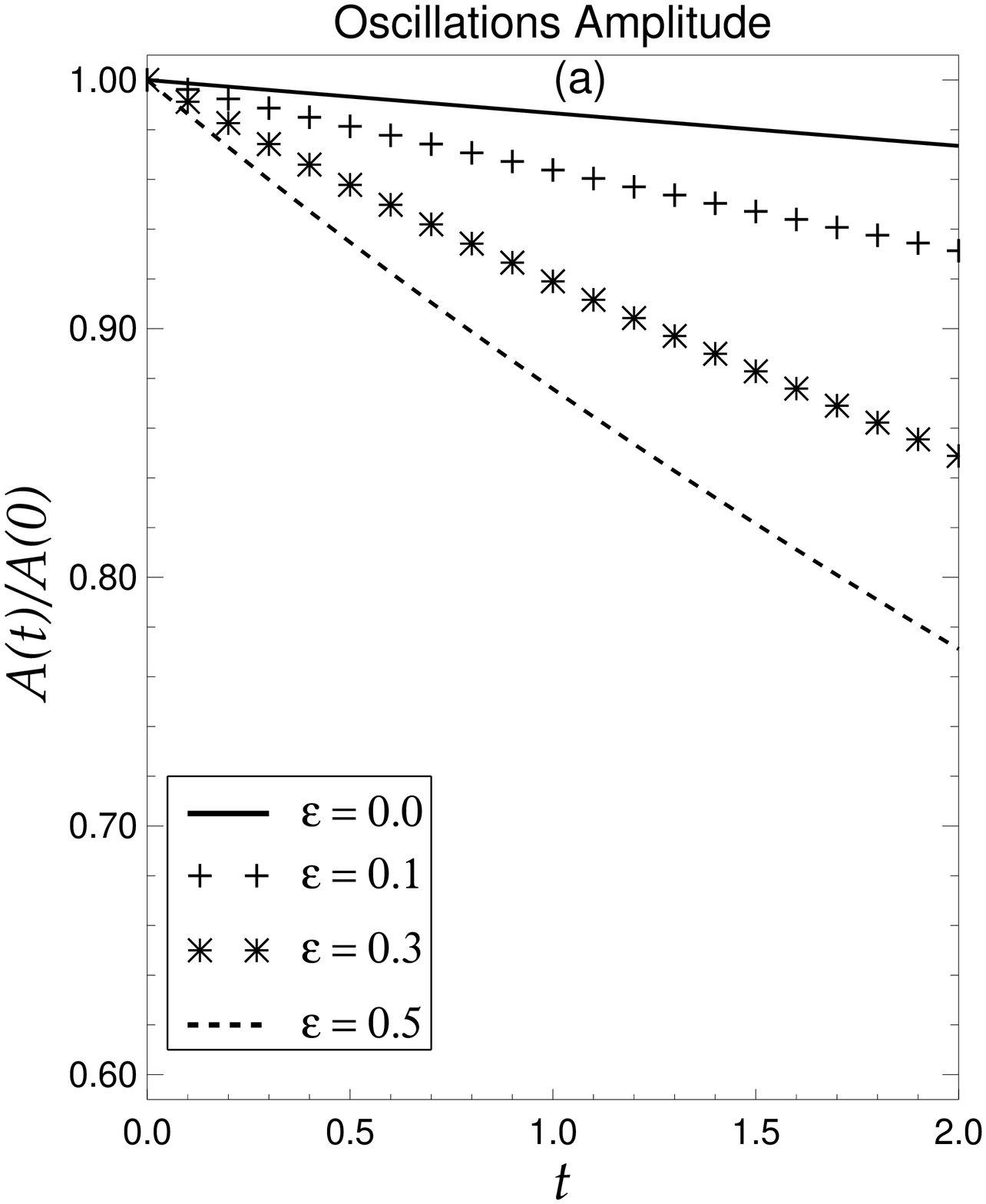}
               \hspace*{-0.03\textwidth}
               \includegraphics[width=0.515\textwidth,clip=]{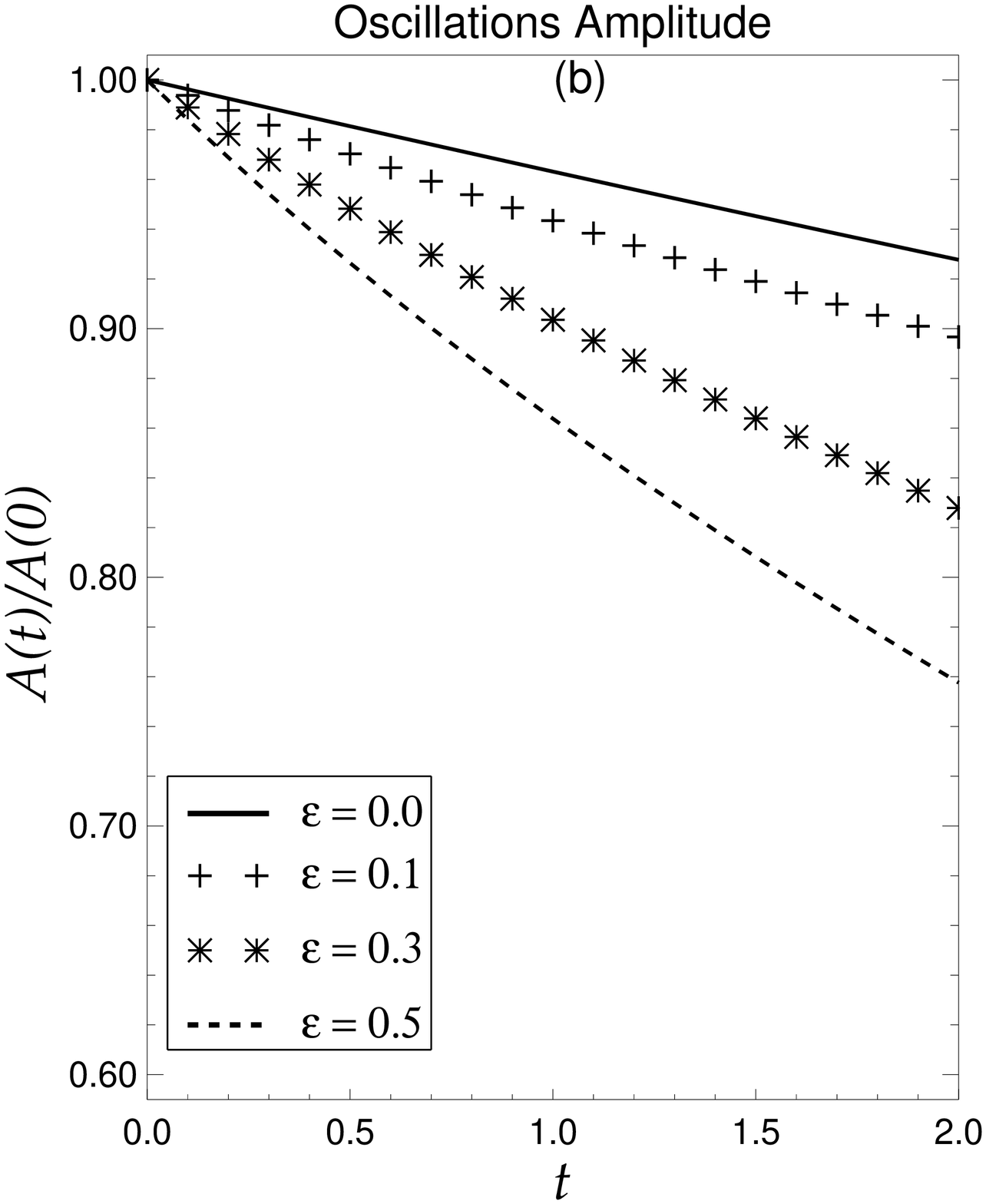}
              }
     \vspace{-0.1\textwidth}   
  
   \centerline{\hspace*{0.015\textwidth}
               \includegraphics[width=0.515\textwidth,clip=]{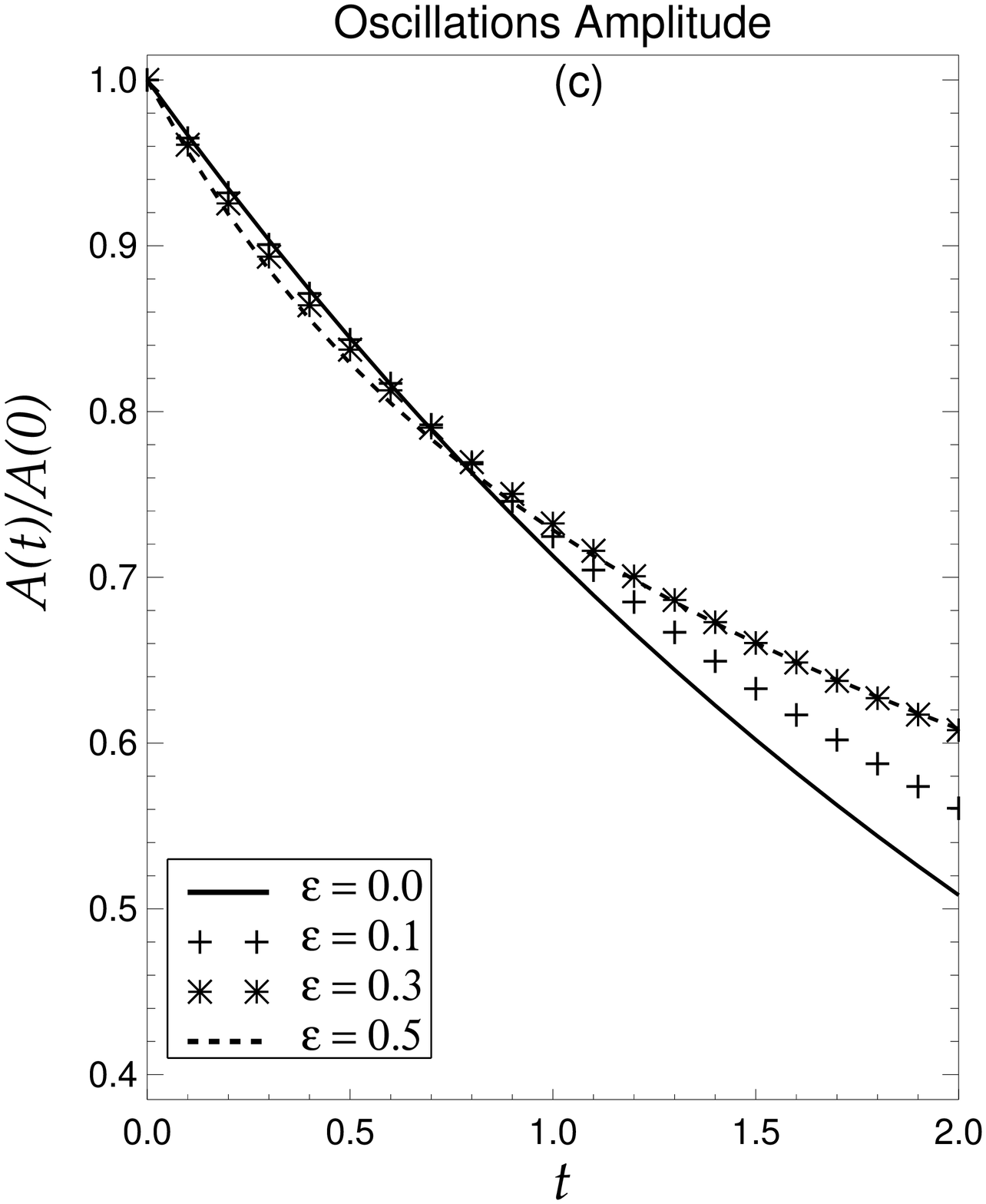}
               \hspace*{-0.03\textwidth}
               \includegraphics[width=0.515\textwidth,clip=]{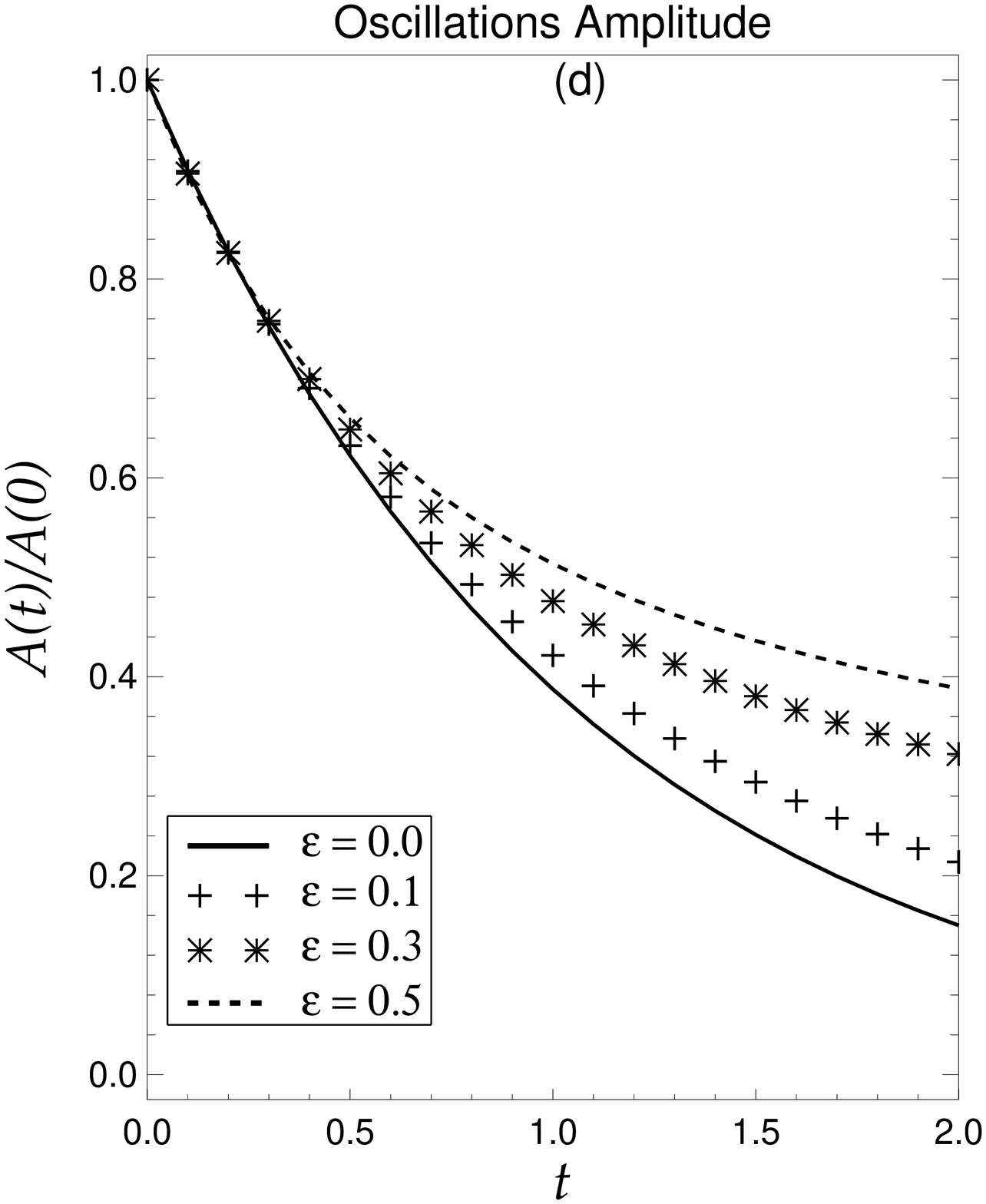}
              }
    \vspace{-0.05\textwidth}   

\caption{The dependence of the oscillation amplitude on time. Panels (a), (b), (c) and (d) correspond to $T_{0i} = 0.6$~MK ($\sigma = 0.0068$), $T_{0i} = 1$~MK ($\sigma=0.019$), $T_{0i} = 3$~MK ($\sigma = 0.17$) and $T_{0i} = 5$~MK ($\sigma = 0.48$), respectively. The time is measured in units of $L/c_{si}$.}
   \label{Amp-damping}
   \end{figure}

The temporal evolution of longitudinal standing-mode amplitude for various values of $\epsilon$ and the initial loop temperature is exhibited in Figure~\ref{Amp-damping}. We can see a remarkable changes in the wave amplitude for different temperature regions. For example, Figures~\ref{Amp-damping}a and \ref{Amp-damping}b indicate that the emergence of cooling enhances the rate of damping caused by thermal conduction for loops with temperature $T_0\le1$ MK. On the other hand, the damping rate of wave amplitude starts to decrease due to the plasma cooling when the loop temperature approaches $3$ MK onwards and the reduction in damping grows quickly by increasing the amount of cooling. However, the state of damping is still rising strongly with time in the absence of cooling as displayed in Figures~\ref{Amp-damping}c and \ref{Amp-damping}d. 

Overall, the scenario of damping in loops with temperature more than 3 MK is in agreement with that obtained by \cite{Al-Ghafri14}
\section{Discussion and Conclusion}\label{conclusion}
In this paper we have investigated the damping of standing longitudinal MHD waves due to radiation and thermal conduction in cooling coronal loops. The plasma cooling is assumed because of radiation method. The radiative function is postulated to have the form $\delta p$ to match the observed cooling which has approximately an exponential profile. Subsequently, the temperature in the loop decreases exponentially with the characteristic time scale which is much longer than the characteristic oscillation period. The assumption of the low-beta plasma reduces the MHD equations to one-dimensional system for standing acoustic waves.  We have used the WKB theory to obtain an analytic solution for the governing MHD equations. 

Typical coronal conditions are applied to evaluate the evolution of amplitude with time numerically. The results show that the radiative cooling enhances the damping rate of coronal loops with temperature $T_0\le1$ MK while in the region of temperature $T_0\ge3$ MK the damping is reduced gradually by cooling. In comparison with radiation mechanism, it is found that thermal conduction is not sufficient to cause a strong damping for very cool loops.  However, the damping of hot coronal loops are mainly dominated by thermal conduction, where the amplitude of standing slow MHD waves experiences a rapid damping with time in the lack of cooling.

According to the results, the rate of damping of coronal oscillations seems to increase continuously with time until reaching its maximum in the region of temperature $4\lesssim T \lesssim6$ MK, and then decreases onwards. Eventually, slow magnetosonic waves will propagate almost without damping when thermal conduction approaches infinity.

%

%
 \begin{acks}
K.S. Al-Ghafri acknowledges the support by College of Applied Sciences - Ibri, (Ministry of Higher Education), Oman.
 \end{acks}

%
%

\begin{thebibliography}{24}
\ifx\bisbn     \undefined \def\bisbn  #1{ISBN #1}\fi
\ifx\binits    \undefined \def\binits#1{#1}\fi
\ifx\bauthor   \undefined \def\bauthor#1{#1}\fi
\ifx\batitle   \undefined \def\batitle#1{#1}\fi
\ifx\bjtitle   \undefined \def\bjtitle#1{\textit{#1}}\fi
\ifx\bvolume   \undefined \def\bvolume#1{\textbf{#1}}\fi
\ifx\byear     \undefined \def\byear#1{#1}\fi
\ifx\bissue    \undefined \def\bissue#1{#1}\fi
\ifx\bfpage    \undefined \def\bfpage#1{#1}\fi
\ifx\blpage    \undefined \def\blpage #1{#1}\fi
\ifx\burl      \undefined \def\burl#1{\textsf{#1}}\fi
\ifx\href      \undefined \def\href#1#2{\textsf{#2}}\fi
\ifx\betal     \undefined \def\betal{\textit{et al.}}\fi
\ifx\bctitle   \undefined \def\bctitle#1{#1}\fi
\ifx\beditor   \undefined \def\beditor#1{#1}\fi
\ifx\bbtitle   \undefined \def\bbtitle#1{\textit{#1}}\fi
\ifx\bedition  \undefined \def\bedition#1{#1}\fi
\ifx\bseriesno \undefined \def\bseriesno#1{\textbf{#1}}\fi
\ifx\blocation \undefined \def\blocation#1{#1}\fi
\ifx\bsertitle \undefined \def\bsertitle#1{\textit{#1}}\fi
\ifx\bsnm      \undefined \def\bsnm#1{#1}\fi
\ifx\bsuffix   \undefined \def\bsuffix#1{#1}\fi
\ifx\bparticle \undefined \def\bparticle#1{#1}\fi
\ifx\barticle  \undefined \def\barticle#1{}\fi
\ifx\binstitute  \undefined \def\binstitute#1{#1}\fi
\ifx\bpublisher  \undefined \def\bpublisher#1{#1}\fi
\ifx\doiurl    \undefined
  \def\doiurl#1{\href{http://dx.doi.org/#1}{\textsf{DOI}}}\fi
\ifx\arxivurl  \undefined
  \def\arxivurl#1{\href{http://arxiv.org/abs/#1}{\textsf{arXiv}}}\fi
\ifx\adsurl    \undefined
  \def\adsurl#1{\href{http://adsabs.harvard.edu/abs/#1}{\textsf{ADS}}}\fi
\ifx\botherref \undefined \def\botherref#1{}\fi
\ifx\url       \undefined \def\url#1{\textsf{#1}}\fi
\ifx\bchapter  \undefined \def\bchapter#1{}\fi
\ifx\bbook     \undefined \def\bbook#1{}\fi
\ifx\bcomment  \undefined \def\bcomment#1{#1}\fi
\ifx\oauthor   \undefined \def\oauthor#1{#1}\fi
\ifx\citeauthoryear \undefined\def \citeauthoryear#1{#1}\fi
\def\endbibitem {}
\ifx\bconflocation  \undefined \def\bconflocation#1{#1} \fi

\bibitem[\protect\citeauthoryear{Al-Ghafri and Erd\'{e}lyi}{2013}]{Al-Ghafri12}
\begin{barticle}
\bauthor{\bsnm{Al-Ghafri}, \binits{K.S.}},
\bauthor{\bsnm{Erd\'{e}lyi}, \binits{R.}}:
\byear{2013},
\bjtitle{Solar Phys.}
\bvolume{283},
\bfpage{413}.
\end{barticle}
\endbibitem

\bibitem[\protect\citeauthoryear{Al-Ghafri \textit{et~al.}}{2014}]{Al-Ghafri14}
\begin{barticle}
\bauthor{\bsnm{Al-Ghafri}, \binits{K.S.}},
\bauthor{\bsnm{Ruderman}, \binits{M.S.}},
\bauthor{\bsnm{Williamson}, \binits{A.}},
\bauthor{\bsnm{Erd\'{e}lyi}, \binits{R.}}:
\byear{2014},
\bjtitle{Astrophys. J.}
\bvolume{786},
\bfpage{36}.
\end{barticle}
\endbibitem

\bibitem[\protect\citeauthoryear{{Aschwanden} and
  {Terradas}}{2008}]{Aschwanden08}
\begin{barticle}
\bauthor{\bsnm{{Aschwanden}}, \binits{M.J.}},
\bauthor{\bsnm{{Terradas}}, \binits{J.}}:
\byear{2008},
\bjtitle{Astrophys. J.}
\bvolume{686},
\bfpage{L127}.
\end{barticle}
\endbibitem

\bibitem[\protect\citeauthoryear{{De Moortel} and {Hood}}{2003}]{Moortel03}
\begin{barticle}
\bauthor{\bsnm{{De Moortel}}, \binits{I.}},
\bauthor{\bsnm{{Hood}}, \binits{A.W.}}:
\byear{2003},
\bjtitle{Astron. Astrophys.}
\bvolume{408},
\bfpage{755}.
\end{barticle}
\endbibitem

\bibitem[\protect\citeauthoryear{Erd\'{e}lyi, Al-Ghafri, and
  Morton}{2011}]{Erdelyi11}
\begin{barticle}
\bauthor{\bsnm{Erd\'{e}lyi}, \binits{R.}},
\bauthor{\bsnm{Al-Ghafri}, \binits{K.S.}},
\bauthor{\bsnm{Morton}, \binits{R.J.}}:
\byear{2011},
\bjtitle{Solar Phys.}
\bvolume{272},
\bfpage{73}.
\end{barticle}
\endbibitem

\bibitem[\protect\citeauthoryear{Kliem \textit{et~al.}}{2002}]{Kliem02}
\begin{barticle}
\bauthor{\bsnm{Kliem}, \binits{B.}},
\bauthor{\bsnm{Dammasch}, \binits{I.E.}},
\bauthor{\bsnm{Curdt}, \binits{W.}},
\bauthor{\bsnm{Wilhelm}, \binits{K.}}:
\byear{2002},
\bjtitle{Astrophys. J. Lett.}
\bvolume{568},
\bfpage{L61}.
\end{barticle}
\endbibitem

\bibitem[\protect\citeauthoryear{Klimchuk}{2012}]{Klimchuk12}
\begin{barticle}
\bauthor{\bsnm{Klimchuk}, \binits{J.A.}}:
\byear{2012},
\bjtitle{J. Geophys. Res.}
\bvolume{117},
\bfpage{A12102}.
\end{barticle}
\endbibitem

\bibitem[\protect\citeauthoryear{{Mendoza-Brice\~{n}o}, {Erd\'{e}lyi}, and
  {Sigalotti}}{2004}]{Mendoza04}
\begin{barticle}
\bauthor{\bsnm{{Mendoza-Brice\~{n}o}}, \binits{C.A.}},
\bauthor{\bsnm{{Erd\'{e}lyi}}, \binits{R.}},
\bauthor{\bsnm{{Sigalotti}}, \binits{L.D.G.}}:
\byear{2004},
\bjtitle{Astrophys. J.}
\bvolume{605},
\bfpage{493}.
\end{barticle}
\endbibitem

\bibitem[\protect\citeauthoryear{{Morton} and {Erd\'{e}lyi}}{2009}]{Morton09b}
\begin{barticle}
\bauthor{\bsnm{{Morton}}, \binits{R.}},
\bauthor{\bsnm{{Erd\'{e}lyi}}, \binits{R.}}:
\byear{2009},
\bjtitle{Astrophys. J.}
\bvolume{707},
\bfpage{750}.
\end{barticle}
\endbibitem

\bibitem[\protect\citeauthoryear{{Morton} and {Erd\'{e}lyi}}{2010}]{Morton09c}
\begin{barticle}
\bauthor{\bsnm{{Morton}}, \binits{R.}},
\bauthor{\bsnm{{Erd\'{e}lyi}}, \binits{R.}}:
\byear{2010},
\bjtitle{Astrophys. J.}
\bvolume{519},
\bfpage{A43}.
\end{barticle}
\endbibitem

\bibitem[\protect\citeauthoryear{{Morton}, {Hood}, and
  {Erd\'{e}lyi}}{2010}]{Morton10a}
\begin{barticle}
\bauthor{\bsnm{{Morton}}, \binits{R.}},
\bauthor{\bsnm{{Hood}}, \binits{A.W.}},
\bauthor{\bsnm{{Erd\'{e}lyi}}, \binits{R.}}:
\byear{2010},
\bjtitle{Astrophys. J.}
\bvolume{512},
\bfpage{A23}.
\end{barticle}
\endbibitem

\bibitem[\protect\citeauthoryear{{Ofman} and {Wang}}{2002}]{Ofmanwang}
\begin{barticle}
\bauthor{\bsnm{{Ofman}}, \binits{L.}},
\bauthor{\bsnm{{Wang}}, \binits{T.}}:
\byear{2002},
\bjtitle{Astrophys. J.}
\bvolume{580},
\bfpage{L85}.
\end{barticle}
\endbibitem

\bibitem[\protect\citeauthoryear{{Priest}}{2000}]{Priest}
\begin{bbook}
\bauthor{\bsnm{{Priest}}, \binits{E.R.}}:
\byear{2000},
\bbtitle{Solar magneto-hydrodynamics},
\bpublisher{Kluwer Dordrecht}.
\end{bbook}
\endbibitem

\bibitem[\protect\citeauthoryear{Rosner, Tucker, and Vaiana}{1978}]{Rosner78}
\begin{barticle}
\bauthor{\bsnm{Rosner}, \binits{R.}},
\bauthor{\bsnm{Tucker}, \binits{W.H.}},
\bauthor{\bsnm{Vaiana}, \binits{G.S.}}:
\byear{1978},
\bjtitle{Astrophys. J.}
\bvolume{220},
\bfpage{643}.
\end{barticle}
\endbibitem

\bibitem[\protect\citeauthoryear{Ruderman}{2011a}]{Ruderman11a}
\begin{barticle}
\bauthor{\bsnm{Ruderman}, \binits{M.S.}}:
\byear{2011}a,
\bjtitle{Solar Phys.}
\bvolume{271},
\bfpage{41}.
\end{barticle}
\endbibitem

\bibitem[\protect\citeauthoryear{Ruderman}{2011b}]{Ruderman11b}
\begin{barticle}
\bauthor{\bsnm{Ruderman}, \binits{M.S.}}:
\byear{2011}b,
\bjtitle{Astron. Astrophys.}
\bvolume{534},
\bfpage{A78}.
\end{barticle}
\endbibitem

\bibitem[\protect\citeauthoryear{Sigalotti, Mendoza-Brice\~{n}o, and
  Luna-Cardozo}{2007}]{Sigalotti07}
\begin{barticle}
\bauthor{\bsnm{Sigalotti}, \binits{L.D.G.}},
\bauthor{\bsnm{Mendoza-Brice\~{n}o}, \binits{C.A.}},
\bauthor{\bsnm{Luna-Cardozo}, \binits{M.}}:
\byear{2007},
\bjtitle{Solar Phys.}
\bvolume{246},
\bfpage{187}.
\end{barticle}
\endbibitem

\bibitem[\protect\citeauthoryear{Taroyan \textit{et~al.}}{2005}]{Taroyan05}
\begin{barticle}
\bauthor{\bsnm{Taroyan}, \binits{Y.}},
\bauthor{\bsnm{Erd\'{e}lyi}, \binits{R.}},
\bauthor{\bsnm{Doyle}, \binits{J.G.}},
\bauthor{\bsnm{Bradshaw}, \binits{S.J.}}:
\byear{2005},
\bjtitle{Astron. Astrophys.}
\bvolume{438},
\bfpage{713}.
\end{barticle}
\endbibitem

\bibitem[\protect\citeauthoryear{{Taroyan} \textit{et~al.}}{2007}]{Taroyan07}
\begin{barticle}
\bauthor{\bsnm{{Taroyan}}, \binits{Y.}},
\bauthor{\bsnm{{Erd\'{e}lyi}}, \binits{R.}},
\bauthor{\bsnm{{Wang}}, \binits{T.J.}},
\bauthor{\bsnm{{Bradshaw}}, \binits{S.J.}}:
\byear{2007},
\bjtitle{Astrophys. J.}
\bvolume{659},
\bfpage{L173}.
\end{barticle}
\endbibitem

\bibitem[\protect\citeauthoryear{Viall and Klimchuk}{2012}]{viall12}
\begin{barticle}
\bauthor{\bsnm{Viall}, \binits{N.M.}},
\bauthor{\bsnm{Klimchuk}, \binits{J.A.}}:
\byear{2012},
\bjtitle{Astrophys. J.}
\bvolume{753},
\bfpage{35}.
\end{barticle}
\endbibitem

\bibitem[\protect\citeauthoryear{Viall and Klimchuk}{2013}]{viall13}
\begin{barticle}
\bauthor{\bsnm{Viall}, \binits{N.M.}},
\bauthor{\bsnm{Klimchuk}, \binits{J.A.}}:
\byear{2013},
\bjtitle{Astrophys. J.}
\bvolume{771},
\bfpage{115}.
\end{barticle}
\endbibitem

\bibitem[\protect\citeauthoryear{Wang}{2011}]{Wang11}
\begin{barticle}
\bauthor{\bsnm{Wang}, \binits{T.}}:
\byear{2011},
\bjtitle{Space Sci. Rev.}
\bvolume{158},
\bfpage{397}.
\end{barticle}
\endbibitem

\bibitem[\protect\citeauthoryear{{Wang} \textit{et~al.}}{2002}]{Wang02}
\begin{barticle}
\bauthor{\bsnm{{Wang}}, \binits{T.J.}},
\bauthor{\bsnm{{Solanki}}, \binits{S.K.}},
\bauthor{\bsnm{{Curdt}}, \binits{W.}},
\bauthor{\bsnm{{Innes}}, \binits{D.E.}},
\bauthor{\bsnm{{Dammasch}}, \binits{I.E.}}:
\byear{2002},
\bjtitle{Astrophys. J.}
\bvolume{574},
\bfpage{L101}.
\end{barticle}
\endbibitem

\bibitem[\protect\citeauthoryear{{Wang} \textit{et~al.}}{2003}]{Wang03b}
\begin{barticle}
\bauthor{\bsnm{{Wang}}, \binits{T.J.}},
\bauthor{\bsnm{{Solanki}}, \binits{S.K.}},
\bauthor{\bsnm{{Innes}}, \binits{D.E.}},
\bauthor{\bsnm{{Curdt}}, \binits{W.}},
\bauthor{\bsnm{{Marsch}}, \binits{E.}}:
\byear{2003},
\bjtitle{Astrophys. J.}
\bvolume{402},
\bfpage{L17}.
\end{barticle}
\endbibitem

\end{thebibliography}

\end{article} 
\end{document}